\documentclass[twocolumn,aps,pra]{revtex4-1}

\usepackage[normalem]{ulem}
\usepackage{epsfig,pstricks}

\usepackage{amsmath,amssymb}
\DeclareTextSymbolDefault{\i}{OT1} 
\def\NN{{\rm I\kern-.25em}}
\def\P{{\rm I\kern-.25em P}}
\def\ZZ{{\sf Z\kern-.44em Z}}
\def\CC{{\rm\kern.24em \vrule width.04em height1.46ex depth-.07ex
\kern-.30em C}}

\def\RR{{\rm
         \vrule width.04em height1.58ex depth-.0ex
         \kern-.04em R}}
\def\id{{\rm 1\kern-.22em l}}
\def\DJo{$\;$\kern-.4em \hbox{D\kern-.8em\raise.15ex\hbox{--}\kern.35em okovi\'c}}
\newcommand{\beq}{\begin{equation}}
\newcommand{\beqa}{\begin{eqnarray}}
\newcommand{\eeq}{\end{equation}}
\newcommand{\eeqa}{\end{eqnarray}}
\newcommand{\bra}[1]{\langle #1 |}
\newcommand{\ket}[1]{| #1 \rangle}
\newcommand{\diag}{{\rm diag}\;}
\newcommand{\Matrix}[2]{\left( \begin{array}{#1} #2 \end{array}
  \right)}
\newcommand{\eps}{\varepsilon}
\newcommand{\nbeqa}{\begin{eqnarray*}}
\newcommand{\neeqa}{\end{eqnarray*}}
\newcommand{\expect}[1]{\langle #1 \rangle}
\newcommand{\fat}[1]{\mbox{\boldmath $ #1 $\unboldmath}}
\newcommand{\braket}[2]{\langle #1 | #2 \rangle}

\begin{document}

\title{Invariant-based entanglement monotones as expectation values and their 
       experimental detection}
\author{Andreas Osterloh}
\affiliation{Fakult{\"a}t f{\"u}r Physik, Campus Duisburg, 
Universit{\"at} Duisburg-Essen, Lotharstr. 1, 47048 Duisburg, Germany}
\author{Jens Siewert}
\affiliation{Departamento de Qu\'{\i}mica F\'{\i}sica, Universidad del Pa\'{\i}s Vasco UPV/EHU, Apdo. 644, 48080 Bilbao, Spain}
\affiliation{Ikerbasque, Basque Foundation for Science, Alameda Urquijo 36, 48011 Bilbao, Spain}

\begin{abstract}
Characterization and quantification of multipartite entanglement
is one of the challenges in state-of-the-art experiments in quantum information
processing. 
According to theory, this is achieved via entanglement monotones,
that is, functions that do not increase under stochastic local operations 
and classical communication (SLOCC). Typically such monotones include
the wave function and its time-reversal (antilinear-operator formalism)
or they are based on not completely positive maps (e.g., partial transpose). 
Therefore, they are not directly accessible to experimental observations. 
We show how entanglement monotones derived from polynomial local 
SL$(2,\CC)$ invariants
can be re-written in terms of expectation values of observables.
Consequently, the amount of entanglement---of specific SLOCC 
classes---in a given state can be extracted 
from the measurement of correlation 
 functions of local operators.
\end{abstract}

\maketitle

\section{Introduction}
%
The preparation of multipartite quantum states and the characterization 
of their entanglement properties is one of the central issues of 
ongoing experimental research in quantum-information processing. 
Technological progress has made it
possible to generate, e.g., Greenberger-Horne-Zeilinger (GHZ) and $W$ states 
for three and more qubits with
trapped ions (up to fourteen qubits, \cite{Monz2011}) 
and photons
 (up to eight qubits~\cite{Pan2011-8photons}). 
Moreover, an entangled  state that is locally inequivalent 
to the GHZ state, the so-called cluster state could be generated 
with up to eight photons~\cite{Pan2012-cluster}.

The experimental detection of entanglement is a rather subtle
issue~\cite{Leifer04,vanEnk07}.
Until now, the experimental characterization of 
entanglement has been carried out using entanglement 
witnesses or 
Bell inequalities (for a recent review, see Ref.~\cite{GuehneReview2009}). 
The advantage of these methods is that the presence 
of entanglement in a given state can be inferred 
simply from measuring certain correlation functions of local observables. 
Although these methods are not strictly quantitative since to date 
there is no formulation of these concepts such that their results are 
invariant under stochastic local operations and classical 
communication (SLOCC)~\cite{BrussRev}, recent proposals exist on how
(tight) lower bounds can be extracted with variable 
 numerical 
 effort~\cite{Guehne06,Eisert06,OsHyl10,LeeSim2012}.

On the other hand, necessary criteria for the quantification of entanglement 
were worked out and led to the concept of the
{\em entanglement monotone}~\cite{MONOTONES}. Essential
properties are local SU$(2)$ invariance, convexity on the space
of density matrices and non-increasing value under local operations and
classical communication (LOCC). The non-increasing property needs to be
extended to include also stochastic local operations
and classical communication (SLOCC)~\cite{Duer00,SLOCC}. 
The connection between SLOCC and invariance under local SL$(2,\CC)$ was 
established~\cite{Duer00,SLOCC,VerstraeteDM03}.
Based on this, the two SLOCC-inequivalent entanglement classes for three qubits 
were discovered whose representatives
are the $W$ and the GHZ state~\cite{Duer00}. 
While the former state exclusively contains
pairwise entanglement (measured by the concurrence~\cite{Wootters98}), 
a measure for the latter was derived for pure states, 
the three-tangle~\cite{Coffman00}.
Subsequently much work was devoted to the classification of
$N$-partite qubit entanglement 
($N>3$), e.g.,~\cite{VerstraeteDMV02,Miyake03,BriandLT03,OS04,OS05,Akulin06,Lamata07}, 
but up to now it is not clear what the essential criteria for 
such a characterization might be. 
However, it is known that local SL$(2,\CC)$ 
invariance plays a major role~\cite{SL-SU07}, and
SL$(2,\CC)$ invariants---both from classical invariant 
theory~\cite{OLVER,Albeverio01,Jaeger,Luque02,BriandLT03,Luque05}
as well as from other methods~\cite{VerstraeteDM03,OS04}---can 
be written in terms of the coefficients of a state alone 
(and {\em not} their complex conjugate in contrast to SU(2) invariants) and thus 
are naturally 
 represented as expectation values of certain 
antilinear operators~\cite{OS05,DoOs08}. 
This has the drawback that 
those quantities are not directly accessible in an experiment.
Interestingly, also other prominent entanglement detection
methods related to
 not completely positive operators (as the Peres-Horodecki 
criterion~\cite{Peres96,Horodecki96}) 
share this drawback, which might indicate that
sensible measures for entanglement are hard to access 
experimentally for conceptual reasons~\cite{vanEnk06,vanEnk07}.
For measures as the negativity~\cite{Vidal02} the obstacles from dealing 
with not completely
positive maps can be circumvented by adding the identity map~\cite{HoroEkert02}
or by certain contraction techniques~\cite{Horodecki03}.
The addition of an identity map was applied also to the 
concurrence~\cite{HoroPRL03}.

In this work we demonstrate how the antilinear formalism in
Ref.~\cite{OS04,OS05}---and hence {\em all} polynomial
SL$(2,\CC)^{\otimes q}$ invariants~\cite{DoOs08}---can directly 
be rewritten in terms of linear operators
and expectation values thereof by means of an 
invertible mapping~\cite{OstHabil}.
This combines the advantages of experimental accessibility and 
class-specific quantification of entanglement 
(intimately related to SL$(2,\CC)$ invariance).
It provides experimentalists with a tool to
directly measure genuine multipartite qubit entanglement.
The mapping itself does not require qubits as local entities
and, therefore, 
could readily be applied also to SL$(d,\CC)$ invariants
for higher local dimension $d>2$ ({\em e.g.}, for higher spin).
On the other hand, a connection to alternative approaches
directly based on expectation values of linear operators 
(for example, Ref.~\cite{Mintert05}) 
can be established by means of the inverse mapping. 
This comparison also highlights the advantage of an antilinear formulation
as far as the construction of, {\em e.g.}, complete sets of invariants is concerned
for future progress in entanglement classification. 

The paper is organized as follows. After briefly reviewing the antilinear
formalism we present the connection with linear expectation values.
Then we discuss some basic conditions to measure such expectation values
in an experiment.

\section{Antilinear formalism}
%
In Refs.~\cite{OS04,OS05} we have shown that SLOCC
invariants can be obtained via the expectation values of antilinear operators,
as a straightforward generalization of the well-known entanglement measures
concurrence $C(\phi)=\left|\bra{\phi}\sigma_y\otimes\sigma_y {\mathfrak C}\ket{\phi}\right|$ 
for a two-qubit state $\ket{\phi}$
and three-tangle $\tau_3(\chi)$ for a three-qubit state $\ket{\chi}$:
\begin{widetext}
\beqa
\tau_3(\chi)&=&  \left|\bra{\chi}\sigma_\mu\otimes\sigma_y\otimes\sigma_y {\mathfrak C}\ket{\chi}  
          \bra{\chi}\sigma^\mu\otimes\sigma_y\otimes\sigma_y {\mathfrak C}\ket{\chi}\right|\label{tau3} \\
            &=&  \left|\bra{\chi}\bullet \bra{\chi} (\sigma_{\mu}\otimes\sigma_y\otimes\sigma_y)\bullet
                                                         (\sigma^{\mu}\otimes\sigma_y\otimes \sigma_y) 
                                                           {\mathfrak C} 
                            \ket{\chi}\bullet\ket{\chi}
                      \right|\ \ .
\label{3tangle_op}
\eeqa
\end{widetext}
Here the symbol $\bullet$ indicates a tensor product where the $j$th factor
is related to the $j$th copy of $\ket{\chi}$, $\sigma_{\mu}$ are the Pauli
matrices ($\sigma_y\equiv \sigma_2$, $\sigma_0\equiv\id$) and 
${\mathfrak C}$ is the complex conjugation operator.
The ``metric'' for the contraction of lower and upper indices 
is $g_{\mu\nu} = \diag\{1,-1,0,-1\}$.
In Eq.~(\ref{3tangle_op}) it is evident how the three-tangle can be written
as an expectation value of an antilinearly Hermitian operator with respect
to a multiple copy of the state $\ket{\chi}$.
The basic ingredient for the construction of global antilinear operators
(``filters'') are local operators~(``combs'') 
that have zero expectation value for 
any pure state of the local
Hilbert space. The two simplest independent combs are
\beqa
\label{comb1}
{\cal O}_1 &=& \sigma_y {\mathfrak C} \\
{\cal O}_2 &=& \sigma_\mu\bullet\sigma^\mu {\mathfrak C}:=
             \sum_{\mu,\nu=0}^3 g_{\mu\nu}\sigma_\mu\bullet\sigma_\nu {\mathfrak C}\ .
\label{comb2}
\eeqa

\section{Linear expectation values}
%
Consider now a $q$-qubit state $\ket{\psi}$.
Any polynomial $q$-qubit SL$(2,\CC)^{\otimes q}$ invariant 
(an antilinear expectation value of several copies of the 
pure state) can be expressed in terms of
linear observables after rewriting the product of the (complex)
$q$-qubit invariant and its complex conjugate expression
\begin{widetext}
\beqa\nonumber
\bra{\psi}\hat{O}\otimes\dots\ket{\psi^*}&\dots & \bra{\psi^*}
                           \hat{O}^\dagger\otimes\dots\ket{\psi}\dots
=(\bra{\psi}\circ\bra{\psi^*})\bullet\dots
      \left[(\hat{O}\circ\hat{O}^\dagger)\bullet\dots\right]\otimes\dots 
                      (\ket{\psi^*}\circ\ket{\psi})\bullet\dots\\
&=&(\bra{\psi}\circ\bra{\psi^*})\bullet\dots\left[(\hat{O}\circ\hat{O}^\dagger)\bullet\dots\prod_{j}\P_{j,k}\right]\otimes\dots(\ket{\psi}\circ\ket{\psi^*})\bullet\dots      
\eeqa
\end{widetext}
Here we have rearranged the factors and introduced the 
symbol $\circ$ to indicate the tensor product of the $j$th copy
and the corresponding copy from the complex conjugate expression.
The square brackets enclose---for the $k$th qubit---as many pair-wise 
\mbox{$\circ$ products} as there are copies of the state $\ket{\psi}$ in
the antilinear expression of the invariant.
The permutation operator $\P_{j,k}$ exchanges the $j$th copy
of the $k$th qubit with the $j$th copy of the $k$th qubit in the
complex conjugate expression~\footnote{$\P=\frac{1}{2}
                \sigma_\mu\circ\sigma_\mu$, using Einstein summation convention} 
and the product extends (for the $k$th qubit) over all copies $j$ of that
qubit. 

Now, even the complex conjugation in the copies $\ket{\psi^{\ast}}$
can be removed by including a {\em partial} transposition 
acting non-trivially on all operator factors to the right of a $\circ$ symbol. 
It is worth mentioning that since we deal with Hermitian operators,
this partial transpose is equivalent to a {\em partial complex conjugation}.
We quote the result for
 the concurrence
as the simplest illustration of the presented invertible mapping. 
The operator corresponding to $\sigma_y$ is found to be 
$\frac{1}{2}M_{\mu\nu}\sigma_\mu \circ \sigma_\nu$ 
and we can write
\beqa
  |C(\phi)|^2 & =   \bra{\phi}\sigma_y\otimes\sigma_y \ket{\phi^{\ast}}
                          \bra{\phi^{\ast}}\sigma_y\otimes\sigma_y \ket{\phi}
                   \ \ \ \ \ \ \ \ \ \ \ \ \ \ \ \ \ \ \
\label{conc_linear}
\\         =  &  
                    \frac{1}{4}M_{\mu\nu}M_{\kappa\lambda}\bra{\phi}\circ\bra{\phi} (\sigma_{\mu}\otimes\sigma_{\kappa}) \circ
                                        (\sigma_{\nu}\otimes\sigma_{\lambda})\ket{\phi} \circ\ket{\phi}\ 
       \nonumber
\eeqa
where $M_{\mu\nu}=\diag \{1,-1,-1,-1\}$.
In the above equation, the symbol ``$\circ$'' stands for a usual tensor product.
We use a different symbol for highlighting that different copies of the 
same state are multiplicatively combined where the different copies 
arise from taking the absolute square of an SL$(2,\CC)$ invariant.
In contrast, ``$\bullet$'' indicates tensor products of different copies
of the pure state leading to a given homogeneous degree of the (complex)
polynomial SL$(2,\CC)$ invariant (see above), and ``$\otimes$'' 
represents the usual
tensor product structure of the multi-qubit Hilbert space.
Note that the metric $M_{\mu\nu}$ for the summation in Eq.~(\ref{conc_linear}) 
is precisely the Minkowski metric, which is the invariant metric with respect to SL$(2,\CC)$.

Hence, we have found a rule to write the linear operator corresponding 
to a comb 
${\cal O}=\hat{O}\ {\mathfrak C}$ as
\beq\label{mapping}
{\mathfrak L}[\hat{O}]:=(\id \circ{\mathfrak T})[({\hat{O}} \circ{\hat{O}})
               \P]
\eeq
where the partial transposition operator $\mathfrak T$ 
acts only on the operator factors in $\cal{O}$ and $\P$ 
to the right of the $ \circ$ symbol. Moreover, $\P=\prod_j \P_{j,k}$.
For the comb $\sigma_\mu\bullet\sigma^\mu$ we find
\beq
{\mathfrak L}[\sigma_\mu\bullet\sigma^\mu] = {\mathfrak G}_{\kappa\lambda\mu\nu}
(\sigma_\kappa\bullet\sigma_\lambda)\circ (\sigma_\mu\bullet\sigma_\nu) 
\eeq
where we have defined
\beq
{\mathfrak G}_{\kappa\lambda\mu\nu} = \frac{1}{2}
(M_{\kappa\lambda}M_{\mu\nu} - \frac{1}{2} M_{\kappa\mu}M_{\lambda\nu}+ 
            M_{\kappa\nu}M_{\lambda\mu})
\eeq
These results represent a prescription to express 
arbitrary polynomial SL$(2,\CC)^{\otimes q}$ invariants in terms of 
expectation values of linear operators.
The principal idea of the mapping~\eqref{mapping} is generic to
SL$(2,\CC)$ invariants and not limited to qubits.
Clearly, the elegance of the antilinear formalism 
can be noted by comparing Eq.~(\ref{tau3}) to the more 
cumbersome representation of the three-tangle 
in terms of spin correlation functions (the formal existence of such an
expression for the three-tangle in terms of projectors was observed in Ref.~\cite{YuSong2007})
\begin{widetext}
\beqa
|\tau_3(\chi)|^2&=&\frac{1}{16}
M_{\mu_1\mu_3}M_{\mu_2\mu_4}M_{\nu_1\nu_3}M_{\nu_2\nu_4}
{\mathfrak G}_{\kappa\lambda\mu\nu}
\bra{\chi}\sigma_{\mu_1}\sigma_{\mu_2}\sigma_{\kappa}\ket{\chi}
\bra{\chi}\sigma_{\nu_1}\sigma_{\nu_2}\sigma_{\lambda}\ket{\chi}\nonumber\\
&&\ \bra{\chi}\sigma_{\mu_3}\sigma_{\mu_4}\sigma_{\mu}\ket{\chi}
\bra{\chi}\sigma_{\nu_3}\sigma_{\nu_4}\sigma_{\nu}\ket{\chi}\ \ .
\label{tau3_linear}\\
&=& \frac{1}{9}{\mathfrak G}_{\kappa_1\lambda_1\mu_1\nu_1}
{\mathfrak G}_{\kappa_2\lambda_2\mu_2\nu_2}
{\mathfrak G}_{\kappa_3\lambda_3\mu_3\nu_3}
\bra{\chi}\sigma_{\kappa_1}\sigma_{\kappa_2}\sigma_{\kappa_3}\ket{\chi}
\bra{\chi}\sigma_{\lambda_1}\sigma_{\lambda_2}\sigma_{\lambda_3}\ket{\chi}\nonumber\\
&&\ \bra{\chi}\sigma_{\mu_1}\sigma_{\mu_2}\sigma_{\mu_3}\ket{\chi}
\bra{\chi}\sigma_{\nu_1}\sigma_{\nu_2}\sigma_{\nu_3}\ket{\chi}\ \ .
\label{tau3_linear2}
\eeqa
\end{widetext}
The manifest ``Lorentz invariance'' of Eqs.~\eqref{tau3_linear},
\eqref{tau3_linear2} is striking.
While simplicity favors the antilinear framework, 
Eqs.~\eqref{tau3_linear},~\eqref{tau3_linear2} are 
important both from a 
conceptual and a practical point of view:
they indicate how the three-tangle of a pure 
state could be directly measured in an experiment. It is necessary
to emphasize again that the correlation functions in these equations 
(as well as in Eq.~\eqref{conc_linear})
are just the usual {\em one-copy} expectation values that are measured
in standard quantum tomography, as opposed to other
approaches, e.g.,  Ref.~\cite{Mintert05}.

An interesting conclusion from this approach is
that it clarifies the relation between SL$(2,\CC)$ invariants and a proposal
to use the projection operators $P_-$ and $P_+$ to assess entanglement in 
multipartite qubit systems~\cite{Mintert05}.
To this end we note that the linear operator corresponding to 
$\sigma_y{\mathfrak C}$ (cf.~Eq.~(\ref{conc_linear})) equals precisely 
the operator $P_-$ in Ref.~\cite{Mintert05},
which proves SL$(2,\CC)$ invariance of $P_-$.
On the other hand, $P_+$ is not invariant under SL$(2,\CC)$, 
which becomes evident 
on transforming it back to the antilinear framework
\begin{widetext}
\beq
\bra{\psi} \circ\bra{\psi}P_+\ket{\psi} \circ\ket{\psi}=
\gamma_{\mu\nu}\bra{\psi}\sigma_\mu\ket{\psi^*}
\bra{\psi^*}\sigma_\nu\ket{\psi} \ ;\qquad
\gamma_{\mu\nu}= \diag\{1,1,0,1\}\ 
\eeq
\end{widetext}
and take into account the results from Ref.~\cite{OS04}.
As a consequence, an expression for qubits constructed exclusively from $P_-$
in the way proposed in Ref.~\cite{Mintert05}
coincides with the well-known $N$-tangle proposed by Wong and 
Christensen~\cite{Wong00} for an even number of qubits.
Its use as a measure
of multipartite entanglement has certain limitations~\cite{Wong00,ORDERING},
but an analytic convex roof extension is available~\cite{Uhlmann00}.

On the other hand, 
the mapping~\eqref{mapping} of $\hat{O}=\sigma_y$ and
results from~\cite{DoOs08} show how $P_-$ can be used for constructing
polynomial SL$(2,\CC)$ invariants of higher homogeneous  degree.
The resulting procedure is entirely equivalent 
to employing full singlet contractions
with the spinor metric tensor in the antilinear framework
(see e.g.~\cite{VerstraeteDM03}). The latter
method is considerably less efficient
than using the pair of combs \eqref{comb1},\eqref{comb2} proposed in~\cite{OS04}.

\section{Experimental detection} 
%

Now let us discuss the possibility of measuring a quantity such as the 
three-tangle in Eq.~\eqref{tau3_linear}. 
If the state
$\Psi$ under consideration were pure, its entanglement could 
indeed be determined directly by measuring all the
correlation functions that appear 
on the right-hand side (r.h.s.) of Eq.~\eqref{tau3_linear}, i.e., 
by performing full state tomography. 
This is expensive in terms of resources, but it is easily seen
why: No prior knowledge whatsoever about the state $\Psi$ is required, so it has
to be collected in the tomography. It is well-known that the amount of 
entanglement-related information in a state grows exponentially with 
Hilbert-space dimension, therefore it is not surprising that
``measuring an entanglement measure''
cannot be done efficiently\cite{vanEnk06,vanEnk07}.

In an experiment, however, one has to
 deal with mixed states 
$\rho=(1-\varepsilon)\ket{\Psi}\!\bra{\Psi}+\varepsilon\tilde{\rho}$
where $\Psi$ is usually highly entangled, $\varepsilon \ll 1$, and $\tilde{\rho}$ 
is an admixture.
This has two consequences: \\
{\em (i)} The quantity on the
r.h.s.\ of Eq.~(\ref{tau3_linear}) looses its significance for growing
$\varepsilon$, as the experimentally measured correlation functions 
are expectation values for $\rho$ rather than
for $\Psi$: $\langle \ldots \rangle=\mathrm{tr}(\rho\ldots)$.
Therefore, the combination of experimentally measured correlation functions
on the r.h.s.\ in Eq.~\eqref{tau3_linear}
will deviate from the true three-tangle by a correction that depends on the
fidelity of $\Psi$ in $\rho$.\\
{\em (ii)} The true three-tangle $\tau_3(\rho)$ will typically be reduced
{\em beyond the mere loss of weight} of $\Psi$ in $\rho$, 
$\tau_3(\rho)\leq  (1-\eps)^2\tau_3(\Psi)$. 
This is a general property of the
convex-roof extension and cannot easily be assessed, neither experimentally
nor theoretically. Here we present an attempt to estimate a lower bound
of $\tau_3(\rho)$ for small $\eps$.

In case of non-systematic errors,
it should be reasonable, as a first approach, 
to assume this erroneous admixture $\tilde{\rho}$
to be completely uncorrelated or unpolarized. 
For the discussion of unpolarized noise, we rewrite
\beq\label{rho}
\rho=(1-\eps)\ket{\Psi}\bra{\Psi} + \frac{\eps}{d} \id
\eeq
where $d$ is the dimension of the Hilbert space, i.e., 
for $q$ qubits we have
 $d=2^q$.

Let us first analyze the error contribution of type $(i)$.
That is, the correlation functions in Eq.~\eqref{tau3_linear}
are taken with respect to the state~\eqref{rho} instead of
the pure state $\Psi$ as described above.
The identity matrix in Eq.~\eqref{rho} will give a 
contribution only for those terms that exclusively contain
$\sigma_0$ operators for each qubit
since any appearing Pauli matrix
leads to a vanishing trace. Then, the error of type {\em (i)}
is easily estimated as
\beq\label{error-trace-C}
C^2_{\rm exp}=
(1-2\eps)C^2(\Psi)+\frac{\eps}{2} 
\eeq
for the squared concurrence, and
\begin{widetext}
\beq
\tau_3^{\rm exp}=
(1-2\eps)\ \tau_3(\Psi) \ 
+ \ \frac{\eps}{8\tau_3(\psi)} M_{\lambda_1\lambda_2}M_{\mu_1\mu_2}M_{\nu_1\nu_2}
      \expect{\sigma_{\lambda_1}\sigma_{\mu_1}\sigma_{\nu_1}}
      \expect{\sigma_{\lambda_2}\id\id }\expect{\id\ \sigma_{\mu_2}\sigma_{\nu_2}}
\label{error-trace-tau3}
\eeq
\end{widetext}
for the three-tangle. Here, $C^2_{\mathrm{exp}}$ and $\tau_3^{\rm exp}$
are the values of the respective entanglement measure if one simply substitutes
the measured correlation functions with respect to the state~\eqref{rho} 
into Eqs.~\eqref{conc_linear} or \eqref{tau3_linear}, 
respectively. The relations~\eqref{error-trace-C}, \eqref{error-trace-tau3}
show how those values are related to the entanglement measures of the pure
target state $\Psi$, taking into account only errors of type {\em (i)} and 
ignoring those of type {\em (ii)}. It is obvious that this kind of linear
error analysis can be carried out also if the admixture is not as simple 
as in Eq.~\eqref{rho}.
We emphasize that in all the contractions the indices are
local.   Note that we do not use the squared three-tangle from Eq.~\eqref{tau3_linear}, 
because an SL(2,$\CC$) invariant is an entanglement measure only 
if its homogeneous
degree in the wavefunction coefficients does not exceed four~\cite{EBOS2012}.

For clarity, we sketch
the procedure how to obtain the ``measured three-tangle'' of an unknown
and high-fidelity
target state $\Psi$, experimentally represented in a mixed state $\rho$:
a) perform full quantum state tomography on $\rho$, b) determine the value of
$\varepsilon$, that is, the fidelity of $\Psi$, c) use the measured correlation 
functions and the value of $\varepsilon$ in Eqs.~\eqref{error-trace-C}, 
\eqref{error-trace-tau3} to compute the entanglement measure for $\Psi$.

Concerning the error type {\em (ii)}, 
the admixture will affect the multipartite entanglement
in the state beyond the mere loss of weight $\eps$.
This effect is generic when different entanglement classes coexist
(see Ref.~\cite{LOSU}) 
and is reflected in the convex-roof extension, 
which makes this error type considerably more complex.
That is, even if we correctly take into account the deviation of type {\em (i)},
the resulting three-tangle  will be overestimated.
In order to refine the estimate, we discuss optimal decompositions
of weakly mixed states.

Each decomposition vector of the density matrix \eqref{rho}
can be written as follows
\beq\label{opt-decomp}
\ket{\Phi_i}=U_{i,0}\sqrt{1-\eps}\ket{\Psi}
          +\sum_l U_{i,k}\sqrt{\eps p_k} \ket{\varphi_k}
\eeq
where $\tilde{\rho}=\sum_k p_k \ket{\varphi_k}\bra{\varphi_k}$
and $U$ is a block of a unitary matrix, i.e., its column vectors
$\vec{u}_{k}:=(U_{0,k},U_{1,k},\dots)$ are orthonormal, 
in particular $\sum_i |U_{i,k}|^2=1$.
For $\eps=0$ the optimal decomposition is trivially given by
a single column vector with the only non-zero element $U_{1,0}=1$.
Continuity arguments for the decomposition vectors show that
for $\eps\ll 1$ a $\delta=:\Delta\,\eps \ll 1$
exists such that
\beq\label{Uopt}
U=\Matrix{cccc}{
\beta_1 \\
\vdots &  & \fat{\delta_{i,j}}  \\
\beta_m \\
\hrulefill &\hrulefill &\hrulefill &\hrulefill \\
\alpha_1 \sqrt{\Delta \eps}  \\  
\alpha_2 \sqrt{\Delta \eps} & & \fat{\gamma_{i,j}} \\
\vdots}
\eeq
where $\sum_{j=1}^{N-m} |\alpha_j|^2=1$ and without loss of generality
$\beta\in\RR^+$.
Here, $\Delta$ will be of order $1$.
Then,
\nbeqa
\sum_{j=1}^m \beta_j^2 + \Delta \eps &=& 1\\
\sum_{i=1}^m |\delta_{i,j}|^2 + \sum_{i=1}^{N-m} |\gamma_{i,j}|^2 
&=& 1
\neeqa
It is worth noticing that the matrix elements 
$\fat{\delta}_{i,j}$ and $\fat{\gamma}_{i,j}$ 
will also be of order 1
such that $U$ itself will typically not be continuous 
as the length of the optimal decomposition may change.
The convex roof is obtained as the  average three-tangle
in the optimal decomposition
\beq
\tau_3(\rho)=\sum_i \braket{\Phi_i}{\Phi_i}^{-1}
 |\bra{\Phi_i}\Sigma_{\mu 2 2}\ket{\Phi_i^*}
  \bra{\Phi_i}\Sigma^{\mu 2 2}\ket{\Phi_i^*}|\ \ ,
\eeq
where we have used the abbreviation 
$\Sigma_{\mu 22}:=\sigma_{\mu}\otimes\sigma_2\otimes\sigma_2$.
There is one particular case that needs to be distinguished:
namely if $m=1$ and $\delta_{1,j}\equiv 0$. It corresponds to
a situation, where the convex roof close to the state $\Psi$
is affine in a range well beyond $\eps$. 
Then, the situation simplifies a lot
and a lower bound for the entanglement is simply given by 
$\tau_3(\rho)\geq \tau_3(\psi) (1-2 \Delta \eps)$.
An example for this is a mixture of
a GHZ state and an orthogonal $W$ state as studied in~\cite{LOSU,Kennlinie,Jung09},
where $\Delta=\frac{1}{2}\frac{1}{1-p_1}\approx 1.71$
with
$p_1 = \frac{1}{2}+\frac{3}{310}\sqrt{465}\approx 0.71$.
A drawback of this estimate is, however, that for a generic experimental
situation,
it will be difficult to understand whether this scenario applies and 
what the value of $\Delta$ would be.

It is worthwhile recalling~\cite{Jung09,OsHyl10,Eltschka2012} that,
when the GHZ state is mixed with 
equal weights of $W$ and the flipped $W$ state 
the three-tangle vanishes already for $\eps=1/4$.
From this, in Ref.~\cite{OsHyl10} an {\em upper} bound for the
pseudo-pure state Eq.~\eqref{rho} ($\eps=1/4$) was derived: 
$\tau_3(\rho)\le 1/9$.
An analysis of general decompositions along the lines discussed above 
(we do not show the calculation here) 
cannot compete with such ``heuristic methods'', as it leads to 
estimates for the three-tangle of the state in Eq.~\eqref{rho}
which are by far too conservative.
Recently, the convex-roof problem could be solved for $\sqrt{\tau_3}$
of the so-called GHZ-symmetric states~\cite{Siewert2012}. 
The exact result for the pseudo-pure  GHZ state 
is $\sqrt{\tau_3}(\rho(\eps)) =1-c\, \eps$ (with $c\approx 3.284$) 
which extends up to $\eps\approx 0.304$ where $\sqrt{\tau_3}$ vanishes.
For the estimation of $\tau_3(\rho)$ in a generic case,
an in-depth analysis would be desirable in order to understand how 
a sensible estimate can be found.

We emphasize that, in contrast to the error described by Eq.~\eqref{error-trace-tau3}, 
the estimates for error {\em (ii)} 
represent bounds to the true mixed-state tangle of the state $\rho$.
One must not combine both errors
to a total error, since they correspond to different quantities. 
The deviation in formula~\eqref{error-trace-tau3} could be used to experimentally check
to which extent a maximally entangled state has been produced,
and a subsequent distillation procedure could be applied to improve 
fidelity and entanglement.
In contrast, error {\em (ii)} 
addresses the question how much entanglement there
really is contained in a mixed state that has been produced 
in the laboratory.

\section{Conclusions} 
We present a one-to-one mapping which translates entanglement
measures formulated within an antilinear formalism~\cite{OS04,OS05} 
into expressions in terms of linear operators. These linear expressions
have twice the multilinear degree of their more compact originals in 
the antilinear framework and thus an 
approach to entanglement quantification and 
 classification
most naturally appears within the antilinear formulation. 
Thus, from a formal point of view our results favor an antilinear approach 
in order to obtain a deeper understanding of entanglement in multipartite
qubit systems.
This aspect is particularly important since for defining a proper
entanglement hierarchy in multipartite systems each family
of entanglement should be expected to have its own 
measure~\cite{ORDERING}.
On the other hand, the linear framework explicitly
displays the ``Lorentz invariance'' of the entanglement measures 
which was pointed out
before in many occasions (see, e.g., \cite{Verstraete2001,Jaeger2003}) 
and which may turn out advantageous in an alternative formalism 
to describe entanglement theory. 

An important result of this work is an explicit expression
of the three-tangle in terms of spin expectation values.
The transcription of multipartite entanglement measures
given as expectation values of antilinear operators 
in Ref.\cite{OS04,OS05}
is straightforward with the results presented here, 
but it is not the scope of this work to list them explicitly.
The invertible mapping~\eqref{mapping} from the antilinear to the
linear setting is not limited to applications for qubits. As soon
as antilinear expressions for polynomial invariants of qudit systems
are available, also these can be transcribed using the same technique.
The linear counterparts of entanglement 
measures for true multipartite entanglement
are a key for experimentalists to measure genuine multipartite 
entanglement in a class-specific manner, as a complement to 
using entanglement witnesses. 
\acknowledgments
The authors would like to thank C.\ Eltschka, 
P.\ Horodecki, M.\ Ku{\'s}, 
 and A.\ Uhlmann
 for useful discussions. J.S.\ was supported
by Basque Government grant IT-472-10 and by
UPV/EHU under program UFI 11/55.

%

\begin{thebibliography}{10}%
\makeatletter
\providecommand \@ifxundefined [1]{%
 \ifx #1\undefined \expandafter \@firstoftwo
 \else \expandafter \@secondoftwo
\fi
}%
\providecommand \@ifnum [1]{%
 \ifnum #1\expandafter \@firstoftwo
 \else \expandafter \@secondoftwo
\fi
}%
\providecommand \enquote [1]{``#1''}%
\providecommand \bibnamefont  [1]{#1}%
\providecommand \bibfnamefont [1]{#1}%
\providecommand \citenamefont [1]{#1}%
\providecommand\href[0]{\@sanitize\@href}%
\providecommand\@href[1]{\endgroup\@@startlink{#1}\endgroup\@@href}%
\providecommand\@@href[1]{#1\@@endlink}%
\providecommand \@sanitize [0]{\begingroup\catcode`\&12\catcode`\#12\relax}%
\@ifxundefined \pdfoutput {\@firstoftwo}{%
 \@ifnum{\z@=\pdfoutput}{\@firstoftwo}{\@secondoftwo}%
}{%
 \providecommand\@@startlink[1]{\leavevmode}%
 \providecommand\@@endlink[0]{}%
}{%
 \providecommand\@@startlink[1]{%
  \leavevmode
  \pdfstartlink
   attr{/Border[0 0 1 ]/H/I/C[0 1 1]}%
   user{/Subtype/Link/A<</Type/Action/S/URI/URI(#1)>>}%
  \relax
 }%
 \providecommand\@@endlink[0]{\pdfendlink}%
}%
\providecommand \url  [0]{\begingroup\@sanitize \@url }%
\providecommand \@url [1]{\endgroup\@href {#1}{\urlprefix}}%
\providecommand \urlprefix [0]{URL }%
\providecommand \Eprint[0]{\href }%
\@ifxundefined \urlstyle {%
  \providecommand \doi [1]{doi:\discretionary{}{}{}#1}%
}{%
  \providecommand \doi [0]{doi:\discretionary{}{}{}\begingroup
  \urlstyle{rm}\Url }%
}%
\providecommand \doibase [0]{http://dx.doi.org/}%
\providecommand \Doi[1]{\href{\doibase#1}}%
\providecommand \bibAnnote [3]{%
  \BibitemShut{#1}%
  \begin{quotation}\noindent
    \textsc{Key:}\ #2\\\textsc{Annotation:}\ #3%
  \end{quotation}%
}%
\providecommand \bibAnnoteFile [2]{%
  \IfFileExists{#2}{\bibAnnote {#1} {#2} {\input{#2}}}{}%
}%
\providecommand \typeout [0]{\immediate \write \m@ne }%
\providecommand \selectlanguage [0]{\@gobble}%
\providecommand \bibinfo [0]{\@secondoftwo}%
\providecommand \bibfield [0]{\@secondoftwo}%
\providecommand \translation [1]{[#1]}%
\providecommand \BibitemOpen[0]{}%
\providecommand \bibitemStop [0]{}%
\providecommand \bibitemNoStop [0]{.\EOS\space}%
\providecommand \EOS [0]{\spacefactor3000\relax}%
\providecommand \BibitemShut [1]{\csname bibitem#1\endcsname}%
\bibitem{Monz2011}%
  \BibitemOpen
  \bibfield{author}{%
  \bibinfo {author} {\bibfnamefont{T.}~\bibnamefont{Monz}}, \bibinfo {author}
  {\bibfnamefont{P.}~\bibnamefont{Schindler}}, \bibinfo {author}
  {\bibfnamefont{J.~T.}\ \bibnamefont{Barreiro}}, \bibinfo {author}
  {\bibfnamefont{M.}~\bibnamefont{Chwalla}}, \bibinfo {author}
  {\bibfnamefont{D.}~\bibnamefont{Nigg}}, \bibinfo {author}
  {\bibfnamefont{W.~A.}\ \bibnamefont{Coish}}, \bibinfo {author}
  {\bibfnamefont{M.}~\bibnamefont{Harlander}}, \bibinfo {author}
  {\bibfnamefont{W.}~\bibnamefont{Hansel}}, \bibinfo {author}
  {\bibfnamefont{M.}~\bibnamefont{Hennrich}},\ and\ \bibinfo {author}
  {\bibfnamefont{R.}~\bibnamefont{Blatt}},\ }%
  \bibfield{journal}{%
  \bibinfo {journal} {Physical Review Letters}\ }%
  \textbf{\bibinfo {volume} {106}},\ \bibinfo {pages} {130506} (\bibinfo {year}
  {2011})%
  \bibAnnoteFile{NoStop}{Monz2011}%
\bibitem{Pan2011-8photons}%
  \BibitemOpen
  \bibfield{author}{%
  \bibinfo {author} {\bibfnamefont{X.}~\bibnamefont{Yao}}, \bibinfo {author}
  {\bibfnamefont{T.~X.}\ \bibnamefont{Wang}}, \bibinfo {author}
  {\bibfnamefont{P.}~\bibnamefont{Xu}}, \bibinfo {author}
  {\bibfnamefont{H.}~\bibnamefont{Lu}}, \bibinfo {author}
  {\bibfnamefont{G.~S.}\ \bibnamefont{Pan}}, \bibinfo {author}
  {\bibfnamefont{X.~H.}\ \bibnamefont{Bao}}, \bibinfo {author}
  {\bibfnamefont{C.~Z.}\ \bibnamefont{Peng}}, \bibinfo {author}
  {\bibfnamefont{C.~Y.}\ \bibnamefont{Lu}}, \bibinfo {author}
  {\bibfnamefont{Y.~A.}\ \bibnamefont{Chen}},\ and\ \bibinfo {author}
  {\bibfnamefont{J.~W.}\ \bibnamefont{Pan}},\ }%
  \bibfield{journal}{%
  \bibinfo {journal} {Nature Photonics}\ }%
  \textbf{\bibinfo {volume} {6}},\ \bibinfo {pages} {225} (\bibinfo {year}
  {2011})%
  \bibAnnoteFile{NoStop}{Pan2011-8photons}%
\bibitem{Pan2012-cluster}%
  \BibitemOpen
  \bibfield{author}{%
  \bibinfo {author} {\bibfnamefont{X.~C.}\ \bibnamefont{Yao}}, \bibinfo
  {author} {\bibfnamefont{T.~X.}\ \bibnamefont{Wang}}, \bibinfo {author}
  {\bibfnamefont{H.~Z.}\ \bibnamefont{Chen}}, \bibinfo {author}
  {\bibfnamefont{W.~B.}\ \bibnamefont{Gao}}, \bibinfo {author}
  {\bibfnamefont{A.~G.}\ \bibnamefont{Fowler}}, \bibinfo {author}
  {\bibfnamefont{R.}~\bibnamefont{Raussendorf}}, \bibinfo {author}
  {\bibfnamefont{Z.~B.}\ \bibnamefont{Chen}}, \bibinfo {author}
  {\bibfnamefont{N.~L.}\ \bibnamefont{Liu}}, \bibinfo {author}
  {\bibfnamefont{C.~Y.}\ \bibnamefont{Lu}}, \bibinfo {author}
  {\bibfnamefont{Y.~J.}\ \bibnamefont{Deng}}, \bibinfo {author}
  {\bibfnamefont{Y.~A.}\ \bibnamefont{Chen}},\ and\ \bibinfo {author}
  {\bibfnamefont{J.~W.}\ \bibnamefont{Pan}},\ }%
  \bibfield{journal}{%
  \bibinfo {journal} {Nature}\ }%
  \textbf{\bibinfo {volume} {482}},\ \bibinfo {pages} {489} (\bibinfo {year}
  {2012})%
  \bibAnnoteFile{NoStop}{Pan2012-cluster}%
\bibitem{Leifer04}%
  \BibitemOpen
  \bibfield{author}{%
  \bibinfo {author} {\bibfnamefont{M.~S.}\ \bibnamefont{Leifer}}, \bibinfo
  {author} {\bibfnamefont{N.}~\bibnamefont{Linden}},\ and\ \bibinfo {author}
  {\bibfnamefont{A.}~\bibnamefont{Winter}},\ }%
  \bibfield{journal}{%
  \bibinfo {journal} {Phys Rev A}\ }%
  \textbf{\bibinfo {volume} {69}},\ \bibinfo {pages} {052304} (\bibinfo {year}
  {2004})%
  \bibAnnoteFile{NoStop}{Leifer04}%
\bibitem{vanEnk07}%
  \BibitemOpen
  \bibfield{author}{%
  \bibinfo {author} {\bibfnamefont{S.~J.}\ \bibnamefont{{van Enk}}}, \bibinfo
  {author} {\bibfnamefont{N.}~\bibnamefont{L{\"u}tkenhaus}},\ and\ \bibinfo
  {author} {\bibfnamefont{H.}~\bibnamefont{Kimble}},\ }%
  \bibfield{journal}{%
  \bibinfo {journal} {Phys. Rev. A}\ }%
  \textbf{\bibinfo {volume} {75}},\ \bibinfo {pages} {052318} (\bibinfo {year}
  {2007})%
  \bibAnnoteFile{NoStop}{vanEnk07}%
\bibitem{GuehneReview2009}%
  \BibitemOpen
  \bibfield{author}{%
  \bibinfo {author} {\bibfnamefont{O.}~\bibnamefont{G{\"u}hne}}\ and\ \bibinfo
  {author} {\bibfnamefont{G.}~\bibnamefont{Toth}},\ }%
  \bibfield{journal}{%
  \bibinfo {journal} {Physics Reports}\ }%
  \textbf{\bibinfo {volume} {474}},\ \bibinfo {pages} {1} (\bibinfo {year}
  {2009})%
  \bibAnnoteFile{NoStop}{GuehneReview2009}%
\bibitem{BrussRev}%
  \BibitemOpen
  \bibfield{author}{%
  \bibinfo {author} {\bibfnamefont{D.}~\bibnamefont{Bruss}}, \bibinfo {author}
  {\bibfnamefont{J.~I.}\ \bibnamefont{Cirac}}, \bibinfo {author}
  {\bibfnamefont{P.}~\bibnamefont{Horodecki}}, \bibinfo {author}
  {\bibfnamefont{F.}~\bibnamefont{Hulpke}}, \bibinfo {author}
  {\bibfnamefont{B.}~\bibnamefont{Kraus}}, \bibinfo {author}
  {\bibfnamefont{M.}~\bibnamefont{Lewenstein}},\ and\ \bibinfo {author}
  {\bibfnamefont{A.}~\bibnamefont{Sanpera}},\ }%
  \bibfield{journal}{%
  \bibinfo {journal} {J. Mod. Opt.}\ }%
  \textbf{\bibinfo {volume} {49}},\ \bibinfo {pages} {1399} (\bibinfo {year}
  {2002})%
  \bibAnnoteFile{NoStop}{BrussRev}%
\bibitem{Guehne06}%
  \BibitemOpen
  \bibfield{author}{%
  \bibinfo {author} {\bibfnamefont{O.}~\bibnamefont{G{\"u}hne}}, \bibinfo
  {author} {\bibfnamefont{M.}~\bibnamefont{Reimpell}},\ and\ \bibinfo {author}
  {\bibfnamefont{R.~F.}\ \bibnamefont{Werner}},\ }%
  \bibfield{journal}{%
  \bibinfo {journal} {Phys. Rev. Lett.}\ }%
  \textbf{\bibinfo {volume} {98}},\ \bibinfo {pages} {110502} (\bibinfo {year}
  {2007})%
  \bibAnnoteFile{NoStop}{Guehne06}%
\bibitem{Eisert06}%
  \BibitemOpen
  \bibfield{author}{%
  \bibinfo {author} {\bibfnamefont{J.}~\bibnamefont{Eisert}}, \bibinfo {author}
  {\bibfnamefont{F. G. S. L.}~\bibnamefont{Brand\~ao}},\ and\ \bibinfo {author}
  {\bibfnamefont{K. M. R.}~\bibnamefont{Audenaert}},\ }%
  \bibfield{journal}{%
  \bibinfo {journal} {New J. Phys.}\ }%
  \textbf{\bibinfo {volume} {9}},\ \bibinfo {pages} {46} (\bibinfo {year}
  {2007}),\ \bibinfo {note} {quant-ph/0607163}%
  \bibAnnoteFile{NoStop}{Eisert06}%
\bibitem{OsHyl10}%
  \BibitemOpen
  \bibfield{author}{%
  \bibinfo {author} {\bibfnamefont{A.}~\bibnamefont{Osterloh}}\ and\ \bibinfo
  {author} {\bibfnamefont{P.}~\bibnamefont{Hyllus}},\ }%
  \bibfield{journal}{%
  \bibinfo {journal} {Phys. Rev. A}\ }%
  \textbf{\bibinfo {volume} {81}},\ \bibinfo {pages} {022307} (\bibinfo {year}
  {2010})%
  \bibAnnoteFile{NoStop}{OsHyl10}%
\bibitem{LeeSim2012}%
  \BibitemOpen
  \bibfield{author}{%
  \bibinfo {author} {\bibfnamefont{S.-S.~B.}\ \bibnamefont{Lee}}\ and\ \bibinfo
  {author} {\bibfnamefont{H.-S.}\ \bibnamefont{Sim}},\ }%
  \bibfield{journal}{%
  \bibinfo {journal} {Phys. Rev. A}\ }%
  \textbf{\bibinfo {volume} {85}},\ \bibinfo {pages} {022325} (\bibinfo {year}
  {2012})%
  \bibAnnoteFile{NoStop}{LeeSim2012}%
\bibitem{MONOTONES}%
  \BibitemOpen
  \bibfield{author}{%
  \bibinfo {author} {\bibfnamefont{G.}~\bibnamefont{Vidal}},\ }%
  \bibfield{journal}{%
  \bibinfo {journal} {J.Mod.Opt.}\ }%
  \textbf{\bibinfo {volume} {47}},\ \bibinfo {pages} {355} (\bibinfo {year}
  {2000})%
  \bibAnnoteFile{NoStop}{MONOTONES}%
\bibitem{Duer00}%
  \BibitemOpen
  \bibfield{author}{%
  \bibinfo {author} {\bibfnamefont{W.}~\bibnamefont{D\"ur}}, \bibinfo {author}
  {\bibfnamefont{G.}~\bibnamefont{Vidal}},\ and\ \bibinfo {author}
  {\bibfnamefont{J.~I.}\ \bibnamefont{Cirac}},\ }%
  \bibfield{journal}{%
  \bibinfo {journal} {Phys. Rev. A}\ }%
  \textbf{\bibinfo {volume} {62}},\ \bibinfo {pages} {062314} (\bibinfo {year}
  {2000})%
  \bibAnnoteFile{NoStop}{Duer00}%
\bibitem{SLOCC}%
  \BibitemOpen
  \bibfield{author}{%
  \bibinfo {author} {\bibfnamefont{C.~H.}\ \bibnamefont{Bennett}}, \bibinfo
  {author} {\bibfnamefont{S.}~\bibnamefont{Popescu}}, \bibinfo {author}
  {\bibfnamefont{D.}~\bibnamefont{Rohrlich}}, \bibinfo {author}
  {\bibfnamefont{J.~A.}\ \bibnamefont{Smolin}},\ and\ \bibinfo {author}
  {\bibfnamefont{A.~V.}\ \bibnamefont{Thapliyal}},\ }%
  \bibfield{journal}{%
  \bibinfo {journal} {Phys. Rev. A}\ }%
  \textbf{\bibinfo {volume} {63}},\ \bibinfo {pages} {012307} (\bibinfo {year}
  {2000})%
  \bibAnnoteFile{NoStop}{SLOCC}%
\bibitem{VerstraeteDM03}%
  \BibitemOpen
  \bibfield{author}{%
  \bibinfo {author} {\bibfnamefont{F.}~\bibnamefont{Verstraete}}, \bibinfo
  {author} {\bibfnamefont{J.}~\bibnamefont{Dehaene}},\ and\ \bibinfo {author}
  {\bibfnamefont{B.}~\bibnamefont{{De~Moor}}},\ }%
  \bibfield{journal}{%
  \bibinfo {journal} {Phys. Rev. A}\ }%
  \textbf{\bibinfo {volume} {68}},\ \bibinfo {pages} {012103} (\bibinfo {year}
  {2003})%
  \bibAnnoteFile{NoStop}{VerstraeteDM03}%
\bibitem{Wootters98}%
  \BibitemOpen
  \bibfield{author}{%
  \bibinfo {author} {\bibfnamefont{W.~K.}\ \bibnamefont{Wootters}},\ }%
  \bibfield{journal}{%
  \bibinfo {journal} {Phys. Rev. Lett.}\ }%
  \textbf{\bibinfo {volume} {80}},\ \bibinfo {pages} {2245} (\bibinfo {year}
  {1998})%
  \bibAnnoteFile{NoStop}{Wootters98}%
\bibitem{Coffman00}%
  \BibitemOpen
  \bibfield{author}{%
  \bibinfo {author} {\bibfnamefont{V.}~\bibnamefont{Coffman}}, \bibinfo
  {author} {\bibfnamefont{J.}~\bibnamefont{Kundu}},\ and\ \bibinfo {author}
  {\bibfnamefont{W.~K.}\ \bibnamefont{Wootters}},\ }%
  \bibfield{journal}{%
  \bibinfo {journal} {Phys. Rev. A}\ }%
  \textbf{\bibinfo {volume} {61}},\ \bibinfo {pages} {052306} (\bibinfo {year}
  {2000})%
  \bibAnnoteFile{NoStop}{Coffman00}%
\bibitem{VerstraeteDMV02}%
  \BibitemOpen
  \bibfield{author}{%
  \bibinfo {author} {\bibfnamefont{F.}~\bibnamefont{Verstraete}}, \bibinfo
  {author} {\bibfnamefont{J.}~\bibnamefont{Dehaene}}, \bibinfo {author}
  {\bibfnamefont{B.}~\bibnamefont{{De~Moor}}},\ and\ \bibinfo {author}
  {\bibfnamefont{H.}~\bibnamefont{Verschelde}},\ }%
  \bibfield{journal}{%
  \bibinfo {journal} {Phys. Rev. A}\ }%
  \textbf{\bibinfo {volume} {65}},\ \bibinfo {pages} {052112} (\bibinfo {year}
  {2002})%
  \bibAnnoteFile{NoStop}{VerstraeteDMV02}%
\bibitem{Miyake03}%
  \BibitemOpen
  \bibfield{author}{%
  \bibinfo {author} {\bibfnamefont{A.}~\bibnamefont{Miyake}},\ }%
  \bibfield{journal}{%
  \bibinfo {journal} {Phys. Rev. A}\ }%
  \textbf{\bibinfo {volume} {67}},\ \bibinfo {pages} {012108} (\bibinfo {year}
  {2003})%
  \bibAnnoteFile{NoStop}{Miyake03}%
\bibitem{BriandLT03}%
  \BibitemOpen
  \bibfield{author}{%
  \bibinfo {author} {\bibfnamefont{E.}~\bibnamefont{Briand}}, \bibinfo {author}
  {\bibfnamefont{J.-G.}\ \bibnamefont{Luque}},\ and\ \bibinfo {author}
  {\bibfnamefont{J.-Y.}\ \bibnamefont{Thibon}},\ }%
  \bibfield{journal}{%
  \bibinfo {journal} {J. Phys. A}\ }%
  \textbf{\bibinfo {volume} {36}},\ \bibinfo {pages} {9915} (\bibinfo {year}
  {2003})%
  \bibAnnoteFile{NoStop}{BriandLT03}%
\bibitem{OS04}%
  \BibitemOpen
  \bibfield{author}{%
  \bibinfo {author} {\bibfnamefont{A.}~\bibnamefont{Osterloh}}\ and\ \bibinfo
  {author} {\bibfnamefont{J.}~\bibnamefont{Siewert}},\ }%
  \bibfield{journal}{%
  \bibinfo {journal} {Phys. Rev. A}\ }%
  \textbf{\bibinfo {volume} {72}},\ \bibinfo {pages} {012337} (\bibinfo {year}
  {2005})%
  \bibAnnoteFile{NoStop}{OS04}%
\bibitem{OS05}%
  \BibitemOpen
  \bibfield{author}{%
  \bibinfo {author} {\bibfnamefont{A.}~\bibnamefont{Osterloh}}\ and\ \bibinfo
  {author} {\bibfnamefont{J.}~\bibnamefont{Siewert}},\ }%
  \bibfield{journal}{%
  \bibinfo {journal} {Int. J. Quant. Inf.}\ }%
  \textbf{\bibinfo {volume} {4}},\ \bibinfo {pages} {531} (\bibinfo {year}
  {2006})%
  \bibAnnoteFile{NoStop}{OS05}%
\bibitem{Akulin06}%
  \BibitemOpen
  \bibfield{author}{%
  \bibinfo {author} {\bibfnamefont{A.}~\bibnamefont{Mandilara}}, \bibinfo
  {author} {\bibfnamefont{V.~M.}\ \bibnamefont{Akulin}}, \bibinfo {author}
  {\bibfnamefont{A.~V.}\ \bibnamefont{Smilga}},\ and\ \bibinfo {author}
  {\bibfnamefont{L.}~\bibnamefont{Viola}},\ }%
  \bibfield{journal}{%
  \bibinfo {journal} {Phys. Rev. A}\ }%
  \textbf{\bibinfo {volume} {74}},\ \bibinfo {pages} {022331} (\bibinfo {year}
  {2006})%
  \bibAnnoteFile{NoStop}{Akulin06}%
\bibitem{Lamata07}%
  \BibitemOpen
  \bibfield{author}{%
  \bibinfo {author} {\bibfnamefont{L.}~\bibnamefont{Lamata}}, \bibinfo {author}
  {\bibfnamefont{J.}~\bibnamefont{Leon}}, \bibinfo {author}
  {\bibfnamefont{D.}~\bibnamefont{Salgado}},\ and\ \bibinfo {author}
  {\bibfnamefont{E.}~\bibnamefont{Solano}},\ }%
  \bibfield{journal}{%
  \bibinfo {journal} {Phys. Rev. A}\ }%
  \textbf{\bibinfo {volume} {75}},\ \bibinfo {pages} {022318} (\bibinfo {year}
  {2007})%
  \bibAnnoteFile{NoStop}{Lamata07}%
\bibitem{SL-SU07}%
  \BibitemOpen
  \bibfield{author}{%
  \bibinfo {author} {\bibfnamefont{A.}~\bibnamefont{Osterloh}},\ }%
  \bibfield{journal}{%
  \bibinfo {journal} {App. Phys. B}\ }%
  \textbf{\bibinfo {volume} {98}},\ \bibinfo {pages} {609} (\bibinfo {year}
  {2010})%
  \bibAnnoteFile{NoStop}{SL-SU07}%
\bibitem{OLVER}%
  \BibitemOpen
  \bibfield{author}{%
  \bibinfo {author} {\bibfnamefont{P.}~\bibnamefont{Olver}},\ }%
  \emph{\bibinfo {title} {Classical invariant theory}}\ (\bibinfo {publisher}
  {Cambridge University Press, Cambridge},\ \bibinfo {year} {1999})%
  \bibAnnoteFile{NoStop}{OLVER}%
\bibitem{Albeverio01}%
  \BibitemOpen
  \bibfield{author}{%
  \bibinfo {author} {\bibfnamefont{S.}~\bibnamefont{Albeverio}}\ and\ \bibinfo
  {author} {\bibfnamefont{S.-M.}\ \bibnamefont{Fei}},\ }%
  \bibfield{journal}{%
  \bibinfo {journal} {J. Opt. B}\ }%
  \textbf{\bibinfo {volume} {3}},\ \bibinfo {pages} {223} (\bibinfo {year}
  {2001})%
  \bibAnnoteFile{NoStop}{Albeverio01}%
\bibitem{Jaeger}%
  \BibitemOpen
  \bibfield{author}{%
  \bibinfo {author} {\bibfnamefont{G.~S.}\ \bibnamefont{Jaeger}}, \bibinfo
  {author} {\bibfnamefont{M.}~\bibnamefont{Teodorescu-Frumosu}}, \bibinfo
  {author} {\bibfnamefont{A.~V.}\ \bibnamefont{Sergienko}}, \bibinfo {author}
  {\bibfnamefont{B.~E.~A.}\ \bibnamefont{Saleh}},\ and\ \bibinfo {author}
  {\bibfnamefont{M.~C.}\ \bibnamefont{Teich}},\ }%
  \bibfield{journal}{%
  \bibinfo {journal} {Math. Model. Phys. Eng. Cogn. Sci.}\ }%
  \textbf{\bibinfo {volume} {5}},\ \bibinfo {pages} {273}, V\"axj\"o Univ. Press (\bibinfo {year}
  {2003}),\ \bibinfo {note} {quant-ph/0301174}%
  \bibAnnoteFile{NoStop}{Jaeger}%
\bibitem{Luque02}%
  \BibitemOpen
  \bibfield{author}{%
  \bibinfo {author} {\bibfnamefont{J.-G.}\ \bibnamefont{Luque}}\ and\ \bibinfo
  {author} {\bibfnamefont{J.-Y.}\ \bibnamefont{Thibon}},\ }%
  \bibfield{journal}{%
  \bibinfo {journal} {Phys. Rev. A}\ }%
  \textbf{\bibinfo {volume} {67}},\ \bibinfo {pages} {042303} (\bibinfo {year}
  {2003})%
  \bibAnnoteFile{NoStop}{Luque02}%
\bibitem{Luque05}%
  \BibitemOpen
  \bibfield{author}{%
  \bibinfo {author} {\bibfnamefont{J.-G.}\ \bibnamefont{Luque}}\ and\ \bibinfo
  {author} {\bibfnamefont{J.-Y.}\ \bibnamefont{Thibon}},\ }%
  \bibfield{journal}{%
  \bibinfo {journal} {J. Phys. A}\ }%
  \textbf{\bibinfo {volume} {39}},\ \bibinfo {pages} {371} (\bibinfo {year}
  {2005})%
  \bibAnnoteFile{NoStop}{Luque05}%
\bibitem{DoOs08}%
  \BibitemOpen
  \bibfield{author}{%
  \bibinfo {author} {\bibfnamefont{D.~{\v{Z}}.}\ \bibnamefont{\DJo}}\ and\
  \bibinfo {author} {\bibfnamefont{A.}~\bibnamefont{Osterloh}},\ }%
  \bibfield{journal}{%
  \bibinfo {journal} {J. Math. Phys.}\ }%
  \textbf{\bibinfo {volume} {50}},\ \bibinfo {pages} {033509} (\bibinfo {year}
  {2009})%
  \bibAnnoteFile{NoStop}{DoOs08}%
\bibitem{Peres96}%
  \BibitemOpen
  \bibfield{author}{%
  \bibinfo {author} {\bibfnamefont{A.}~\bibnamefont{Peres}},\ }%
  \bibfield{journal}{%
  \bibinfo {journal} {Phys. Rev. Lett.}\ }%
  \textbf{\bibinfo {volume} {77}},\ \bibinfo {pages} {1413} (\bibinfo {year}
  {1996})%
  \bibAnnoteFile{NoStop}{Peres96}%
\bibitem{Horodecki96}%
  \BibitemOpen
  \bibfield{author}{%
  \bibinfo {author} {\bibfnamefont{R.~H.}\ \bibnamefont{M.~Horodecki},
  \bibfnamefont{P.~Horodecki}},\ }%
  \bibfield{journal}{%
  \bibinfo {journal} {Phys. Lett. A}\ }%
  \textbf{\bibinfo {volume} {223}},\ \bibinfo {pages} {1} (\bibinfo {year}
  {1996})%
  \bibAnnoteFile{NoStop}{Horodecki96}%
\bibitem{vanEnk06}%
  \BibitemOpen
  \bibfield{author}{%
  \bibinfo {author} {\bibfnamefont{S.~J.}\ \bibnamefont{{van Enk}}},\ }%
  \enquote{\bibinfo {title} {Can measuring entanglement be easy?}}\  (\bibinfo
  {year} {2006}),\ \bibinfo {note} {arXiv:quant-ph/0606017}%
  \bibAnnoteFile{NoStop}{vanEnk06}%
\bibitem{Vidal02}%
  \BibitemOpen
  \bibfield{author}{%
  \bibinfo {author} {\bibfnamefont{G.}~\bibnamefont{Vidal}}\ and\ \bibinfo
  {author} {\bibfnamefont{R.~F.}\ \bibnamefont{Werner}},\ }%
  \bibfield{journal}{%
  \bibinfo {journal} {Phys. Rev. A}\ }%
  \textbf{\bibinfo {volume} {65}},\ \bibinfo {pages} {032314} (\bibinfo {year}
  {2002})%
  \bibAnnoteFile{NoStop}{Vidal02}%
\bibitem{HoroEkert02}%
  \BibitemOpen
  \bibfield{author}{%
  \bibinfo {author} {\bibfnamefont{P.}~\bibnamefont{Horodecki}}\ and\ \bibinfo
  {author} {\bibfnamefont{A.}~\bibnamefont{Ekert}},\ }%
  \bibfield{journal}{%
  \bibinfo {journal} {Phys. Rev. Lett.}\ }%
  \textbf{\bibinfo {volume} {89}},\ \bibinfo {pages} {127902} (\bibinfo {year}
  {2002})%
  \bibAnnoteFile{NoStop}{HoroEkert02}%
\bibitem{Horodecki03}%
  \BibitemOpen
  \bibfield{author}{%
  \bibinfo {author} {\bibfnamefont{P.}~\bibnamefont{Horodecki}},\ }%
  \bibfield{journal}{%
  \bibinfo {journal} {Phys. Lett. A}\ }%
  \textbf{\bibinfo {volume} {319}},\ \bibinfo {pages} {1} (\bibinfo {year}
  {2003})%
  \bibAnnoteFile{NoStop}{Horodecki03}%
\bibitem{HoroPRL03}%
  \BibitemOpen
  \bibfield{author}{%
  \bibinfo {author} {\bibfnamefont{P.}~\bibnamefont{Horodecki}},\ }%
  \bibfield{journal}{%
  \bibinfo {journal} {Phys. Rev. Lett.}\ }%
  \textbf{\bibinfo {volume} {90}},\ \bibinfo {pages} {167901} (\bibinfo {year}
  {2003})%
  \bibAnnoteFile{NoStop}{HoroPRL03}%
\bibitem{OstHabil}%
  \BibitemOpen
  \bibfield{author}{%
  \bibinfo {author} {\bibfnamefont{A.}~\bibnamefont{Osterloh}},\ }%
  \enquote{\bibinfo {title} {Entanglement and its facets in condensed matter
  systems},}\ \bibinfo {note} {habilitation thesis, Univ. Hannover, Germany (2008); arXiv:0810.1240}%
  \bibAnnoteFile{NoStop}{OstHabil}%
\bibitem{Mintert05}%
  \BibitemOpen
  \bibfield{author}{%
  \bibinfo {author} {\bibfnamefont{F.}~\bibnamefont{Mintert}}, \bibinfo
  {author} {\bibfnamefont{M.}~\bibnamefont{Ku{\'s}}},\ and\ \bibinfo {author}
  {\bibfnamefont{A.}~\bibnamefont{Buchleitner}},\ }%
  \bibfield{journal}{%
  \bibinfo {journal} {Phys. Rev. Lett.}\ }%
  \textbf{\bibinfo {volume} {95}},\ \bibinfo {pages} {260502} (\bibinfo {year}
  {2005})%
  \bibAnnoteFile{NoStop}{Mintert05}%
\bibitem{Note1}%
  \BibitemOpen
  \bibinfo {note} {${\protect \rm I\kern -.25em P}=\protect \frac {1}{2} \sigma
  _\mu \circ \sigma _\mu $, using Einstein summation convention}%
  \bibAnnoteFile{NoStop}{Note1}%
\bibitem{YuSong2007}%
  \BibitemOpen
  \bibfield{author}{%
  \bibinfo {author} {\bibfnamefont{C.-S.}\ \bibnamefont{Yu}}\ and\ \bibinfo
  {author} {\bibfnamefont{H.-S.}\ \bibnamefont{Song}},\ }%
  \bibfield{journal}{%
  \bibinfo {journal} {Phys. Rev. A}\ }%
  \textbf{\bibinfo {volume} {76}},\ \bibinfo {pages} {022324} (\bibinfo {year}
  {2007})%
  \bibAnnoteFile{NoStop}{YuSong2007}%
\bibitem{Wong00}%
  \BibitemOpen
  \bibfield{author}{%
  \bibinfo {author} {\bibfnamefont{A.}~\bibnamefont{Wong}}\ and\ \bibinfo
  {author} {\bibfnamefont{N.}~\bibnamefont{Christensen}},\ }%
  \bibfield{journal}{%
  \bibinfo {journal} {Phys. Rev. A}\ }%
  \textbf{\bibinfo {volume} {63}},\ \bibinfo {pages} {044301} (\bibinfo {year}
  {2001})%
  \bibAnnoteFile{NoStop}{Wong00}%
\bibitem{ORDERING}%
  \BibitemOpen
  \bibfield{author}{%
  \bibinfo {author} {\bibfnamefont{S.}~\bibnamefont{Virmani}}\ and\ \bibinfo
  {author} {\bibfnamefont{M.~B.}\ \bibnamefont{Plenio}},\ }%
  \bibfield{journal}{%
  \bibinfo {journal} {Physics Letters A}\ }%
  \textbf{\bibinfo {volume} {268}},\ \bibinfo {pages} {31} (\bibinfo {year}
  {2000})%
  \bibAnnoteFile{NoStop}{ORDERING}%
\bibitem{Uhlmann00}%
  \BibitemOpen
  \bibfield{author}{%
  \bibinfo {author} {\bibfnamefont{A.}~\bibnamefont{Uhlmann}},\ }%
  \bibfield{journal}{%
  \bibinfo {journal} {Phys. Rev. A}\ }%
  \textbf{\bibinfo {volume} {62}},\ \bibinfo {pages} {032307} (\bibinfo {year}
  {2000})%
  \bibAnnoteFile{NoStop}{Uhlmann00}%
\bibitem{EBOS2012}%
  \BibitemOpen
  \bibfield{author}{%
  \bibinfo {author} {\bibfnamefont{C.}~\bibnamefont{Eltschka}}, \bibinfo
  {author} {\bibfnamefont{T.}~\bibnamefont{Bastin}}, \bibinfo {author}
  {\bibfnamefont{A.}~\bibnamefont{Osterloh}},\ and\ \bibinfo {author}
  {\bibfnamefont{J.}~\bibnamefont{Siewert}},\ }%
  \bibfield{journal}{%
  \bibinfo {journal} {Phys. Rev. A}\ }%
  \textbf{\bibinfo {volume} {85}},\ \bibinfo {pages} {022301} (\bibinfo {year}
  {2012})%
  \bibAnnoteFile{NoStop}{EBOS2012}%
\bibitem{LOSU}%
  \BibitemOpen
  \bibfield{author}{%
  \bibinfo {author} {\bibfnamefont{R.}~\bibnamefont{Lohmayer}}, \bibinfo
  {author} {\bibfnamefont{A.}~\bibnamefont{Osterloh}}, \bibinfo {author}
  {\bibfnamefont{J.}~\bibnamefont{Siewert}},\ and\ \bibinfo {author}
  {\bibfnamefont{A.}~\bibnamefont{Uhlmann}},\ }%
  \bibfield{journal}{%
  \bibinfo {journal} {Phys. Rev. Lett.}\ }%
  \textbf{\bibinfo {volume} {97}},\ \bibinfo {pages} {260502} (\bibinfo {year}
  {2006})%
  \bibAnnoteFile{NoStop}{LOSU}%
\bibitem{Kennlinie}%
  \BibitemOpen
  \bibfield{author}{%
  \bibinfo {author} {\bibfnamefont{A.}~\bibnamefont{Osterloh}}, \bibinfo
  {author} {\bibfnamefont{J.}~\bibnamefont{Siewert}},\ and\ \bibinfo {author}
  {\bibfnamefont{A.}~\bibnamefont{Uhlmann}},\ }%
  \bibfield{journal}{%
  \bibinfo {journal} {Phys. Rev. A}\ }%
  \textbf{\bibinfo {volume} {77}},\ \bibinfo {pages} {032310} (\bibinfo {year}
  {2008})%
  \bibAnnoteFile{NoStop}{Kennlinie}%
\bibitem{Jung09}%
  \BibitemOpen
  \bibfield{author}{%
  \bibinfo {author} {\bibfnamefont{E.}~\bibnamefont{Jung}}, \bibinfo {author}
  {\bibfnamefont{M.-R.}\ \bibnamefont{Hwang}}, \bibinfo {author}
  {\bibfnamefont{D.}~\bibnamefont{Park}},\ and\ \bibinfo {author}
  {\bibfnamefont{J.-W.}\ \bibnamefont{Son}},\ }%
  \bibfield{journal}{%
  \bibinfo {journal} {Phys. Rev. A}\ }%
  \textbf{\bibinfo {volume} {79}},\ \bibinfo {pages} {024306} (\bibinfo {year}
  {2009})%
  \bibAnnoteFile{NoStop}{Jung09}%
\bibitem{Eltschka2012}%
  \BibitemOpen
  \bibfield{author}{%
  \bibinfo {author} {\bibfnamefont{C.}~\bibnamefont{Eltschka}}\ and\ \bibinfo
  {author} {\bibfnamefont{J.}~\bibnamefont{Siewert}},\ }%
  \bibfield{journal}{%
  \bibinfo {journal} {Phys. Rev. Lett.}\ }%
  \textbf{\bibinfo {volume} {108}},\ \bibinfo {pages} {020502} (\bibinfo {year}
  {2012})%
  \bibAnnoteFile{NoStop}{Eltschka2012}%
\bibitem{Siewert2012}%
  \BibitemOpen
  \bibfield{author}{%
  \bibinfo {author} {\bibfnamefont{J.}~\bibnamefont{Siewert}}\ and\ \bibinfo
  {author} {\bibfnamefont{C.}~\bibnamefont{Eltschka}},\ }%
  \bibfield{journal}{%
  \bibinfo {journal} {Phys. Rev. Lett.}}%
  \textbf{\bibinfo {volume} {108}},\ \bibinfo {pages} {230502}    
  (\bibinfo {year} {2012})%
  \bibAnnoteFile{NoStop}{Siewert2012}%
\bibitem{Verstraete2001}%
  \BibitemOpen
  \bibfield{author}{%
  \bibinfo {author} {\bibfnamefont{F.}~\bibnamefont{Verstraete}}, \bibinfo
  {author} {\bibfnamefont{J.}~\bibnamefont{Dehaene}},\ and\ \bibinfo {author}
  {\bibfnamefont{B.}~\bibnamefont{{De~Moor}}},\ }%
  \bibfield{journal}{%
  \bibinfo {journal} {Phys. Rev. A}\ }%
  \textbf{\bibinfo {volume} {64}},\ \bibinfo {pages} {010101} (\bibinfo {year}
  {2001})%
  \bibAnnoteFile{NoStop}{Verstraete2001}%
\bibitem{Jaeger2003}%
  \BibitemOpen
  \bibfield{author}{%
  \bibinfo {author} {\bibfnamefont{M.}~\bibnamefont{Teodorescu-Frumosu}}\ and\
  \bibinfo {author} {\bibfnamefont{G.}~\bibnamefont{Jaeger}},\ }%
  \bibfield{journal}{%
  \bibinfo {journal} {Phys. Rev. A}\ }%
  \textbf{\bibinfo {volume} {67}},\ \bibinfo {pages} {052305} (\bibinfo {year}
  {2003})%
  \bibAnnoteFile{NoStop}{Jaeger2003}%
\end{thebibliography}

\end{document}